\documentclass[12pt,preprint]{emulateapj}

\usepackage{graphicx}
\usepackage{color}

\newcommand{\beq}{\begin{equation}}
\newcommand{\eeq}{\end{equation}}
\newcommand{\beqa}{\begin{eqnarray}}
\newcommand{\eeqa}{\end{eqnarray}}

\begin{document}

\title{Constraining the distribution of dark matter in inner Galaxy with indirect detection signal: the case of tentative $130~ {\rm GeV}$ $\gamma$-ray Line}

\author{\sc Rui-Zhi Yang\altaffilmark{1,2}, Lei Feng\altaffilmark{1}, Xiang Li\altaffilmark{1,2} and Yi-Zhong Fan\altaffilmark{1,*}}
\altaffiltext{1}{Key Laboratory of Dark Matter and Space Astronomy, Purple Mountain Observatory, Chinese Academy of Sciences, Nanjing 210008, China}
\altaffiltext{2}{Graduate University of Chinese Academy of Sciences, Beijing, 100012, China}
\altaffiltext{*}{To whom the correspondence should be sent (email: yzfan@pmo.ac.cn).}

\begin{abstract}
The dark matter distribution in the very inner region of our Galaxy is still in debate. In the N-body simulations a cuspy dark matter halo density profile is favored. Several dissipative baryonic processes however are found to be able to significantly flatten dark matter distribution and a cored dark matter halo density profile is possible. The baryons dominate the gravitational potential in the inner Galaxy, hence a direct constrain on the abundance of the dark matter particles is rather challenging. Recently, a few groups have identified a tentative 130 GeV line signal in the Galactic center, which could be interpreted as the signal of the dark matter annihilation. With current 130 GeV line data and adopting the generalized Navarro-Frenk-White profile of the dark matter halo, for local dark matter density $\rho_0=0.4~{\rm GeV~cm^{-3}}$ and $r_{\rm s}=20$ kpc we obtain a $95\%$ confidence level lower (upper) limit on the inner slope of dark matter density distribution $\alpha = 1.06$ (the cross section of dark matter annihilation into gamma-rays $\langle\sigma v \rangle_{\chi\chi\rightarrow \gamma\gamma}= 1.3\times 10^{-27}~{\rm cm^{3}~s^{-1}}$). { Such a slope is consistent with the results of some N-body simulations, and if the signal is due to dark matter, suggests that baryonic processes may be unimportant.}
\end{abstract}


\setlength{\parindent}{.25in}

\section{Introduction}
In the leading cold dark matter (CDM) model, structure forms hierarchically
bottom-up, with DM collapsing first into small halos, which then accrete normal matter, merge
and eventually give rise to larger halos.  Galaxies are thought
to form out of gas which cools and collapses to the centers
of these DM haloes (e.g., White \& Rees 1978).
As shown in the high-resolution N-body simulations, the density profiles
of CDM halos can be reasonably well described by an universal form, independent of the halo
mass, and the cuspy density profiles
such as Navarro-Frenk-White (NFW, Navarro et al. 1997) and Einasto \citep{einasto} are found to be favored.
The effect of the baryons ignored in most previous simulations, however, are still to be figured out. One
possibility is the so-called ¡°adiabatic contraction¡±, i.e., when the Galaxy formed and the baryons contracted towards the centre,
dark matter particles are pulled inward and  their central
density increases (e.g.,
Blumenthal et al. 1986; Gnedin et al. 2004). The resulting DM profile of the galaxies is expected to be more cuspy.
Contrary to adiabatic contraction, other baryonic processes, such as the gas bulk
motions, possibly supernova-induced in regions of high star
formation activity, and the subsequent energy loss of gas clouds
due to dynamical friction, can transfer energy to the central
dark matter component or/and induce substantial gravitational potential fluctuations
and finally give rise to a subsequent reduction in the central dark matter density (e.g., Navarro et al. 1996; Mo \& Mao 2004;
El-Zant et al. 2001; Mashchenko et al. 2006; Ogiya \& Mori 2011; Maccio et al. 2012; Pontzen \& Governato 2012). The resulting DM profile of the Galaxies is likely cored.

As shown in the recent high-resolution cosmological
hydrodynamical simulations performed by Maccio et al. (2012; see Fig.1 therein) and by Pontzen \& Governato (2012; see Fig.5 therein), with and without the effects of dissipative baryonic processes the inner distribution of dark matter in Milky-Way-like objects are indeed  rather different.
Nevertheless these dark matter profiles only represent the mean of all simulated halos for
a given mass at a given redshift. The scatter with respect to these mean values
arises plausibly due to the different halo formation histories and due to the evolution of
the expanding Universe { (e.g., Navarro et al. 2010; Wu et al. 2013).} The
 average values may be unable to describe
accurately the DM halo of the Milky Way since its formation and evolution may not follow
a prototypical spiral galaxy. In view of such uncertainties,
observational data is highly needed to reliably constrain the inner structure of the Milky Way DM halo.
However, current microlensing and dynamical data can only rule out some extremely cuspy density profiles (Iocco et al. 2011) since in inner Galaxy, the gravitational potential is dominated by the normal matter rather than the dark matter.
It is well established that the prospect of detecting the products of dark matter particle annihilation is
critically sensitive to the structure of the Milky Way DM halo. Moreover, a robust estimate of the cross section of dark matter particles annihilating into normal matter (photons, electrons/positrons and so on) is not possible unless the DM halo profile has been reasonably well determined. In turn, indirect DM gamma-ray searches can be
used to study the distribution of DM in the innermost regions of the Milky Way
halo as long as a positive signal (in particular an unambiguous gamma-ray line) has been identified.
In this work, we adopt the hypothesis that the Galactic $\sim 130$ GeV  line signal identified in the publicly available Fermi-LAT data firstly by \citet{bringman} and \citet{weniger} originated from the annihilation of dark matter. {  Under this hypothesis, we examine which dark matter density profiles are consistent with the signal, and then discuss the implication.}

\section{Constraining the dark matter density profile and $\langle \sigma v \rangle_{\chi\chi\rightarrow \gamma \gamma}$ with the 130 GeV line signal}
\subsection{The 130 GeV line signal}
High energy $\gamma$-ray line is of greatest interest in looking for the
signal of dark matter annihilation. After analyzing
the publicly available Fermi-LAT $\gamma$-ray data,
\cite{bringman} and \cite{weniger} found possible evidence for a
monochromatic $\gamma$-ray line with energy $\sim130$ GeV.
Later independent analyses confirmed such an excess.
The center of the most prominent signal region
is around the Galactic centre Sgr A$^{\star}$ with an offset $\geq 1.2^\circ$ \citep{tempel,sumeng}, which has been thought to be at odds with the DM origin. However, within the DM annihilation scenario such an offset can be interpreted by the limited statistics of the current $\gamma-$ray line signal consisting of only $\sim 14$ photons \citep{Yang2012}. { Furthermore,  it is interesting to note that there is a small wiggle in the electron spectrum of PAMELA/Fermi-LAT at energies $\sim 100~\rm  GeV$, which may be interpreted as being consistent with the 130 GeV line signal (Feng et al. 2013).}

With a typical cuspy DM density profile such as Navarro-Frenk-White (NFW, \citet{nfw}) and \citet{einasto},
annihilation cross section $\langle\sigma v\rangle_{\rm
\chi\chi\rightarrow \gamma\gamma}\sim 2-5\times 10^{-27} {\rm cm^3~s^{-1}}$ is needed to produce the signal data \citep{weniger,tempel}, which however is larger than the upper limit ($\sim 10^{-27} {\rm cm^3~s^{-1}}$) set by the non-detection of 130 GeV line in other regions by a factor of quite a few. This puzzle has been taken to be a piece of evidence against the dark matter origin of the 130 GeV line in the Milky Way center (e.g., Huang et al. 2012). As already mentioned in Sec. 1, in the Galactic center the DM density profile is rather uncertain, it is thus {\it not} confidential to refute the dark matter model just based on the tension between the values or upper-limits of $\langle\sigma v\rangle_{\rm
\chi\chi\rightarrow \gamma\gamma}$ inferred in different regions.

In principle, one may be able to find an ideal region to reliably evaluate $\langle \sigma v \rangle_{\chi\chi\rightarrow \gamma\gamma}$. In such a region the following conditions should be met, including  (i) The dark matter density distribution is reasonably determined by astrophysical observations; (ii) The signal-to-noise is relatively high; (iii) The contribution of the DM substructure is expected to be not dominated. For the Milky Way, the dynamical data plays a role in constraining the dark matter distribution for $r>3$ kpc, where $r$ is the distance to the Galactic center (e.g., Sofue 2012). Observationally there is still no strong evidence for the existence of abundant substructures of the Milky Way dark matter halo. The N-body simulation Aquarius found out that abundant substructures of the Galaxies might present at $r>20$ kpc (Springel et al. 2008). The contribution to the J-factor may be non-ignorable  (i.e., the contribution is about the same as the smooth halo) any longer at $\psi \approx 30^\circ$. Unless for the very cuspy dark matter halo, the signal-to-noise is also acceptable in the region $\psi \gtrsim30^\circ$ \cite[e.g.,][]{bringman}. In view of these facts, the region $\psi \in (20^\circ,~40^\circ)$ excluding $b\leq 10^\circ$ may be a suitable region to constrain/measure $\langle\sigma v\rangle_{\rm \chi\chi \rightarrow \gamma\gamma}$, where $b$ is the Galactic latitude and $\psi$ is the angle to the Galactic centre. As an unbiased constraint on the dark matter density profile, the data in other regions should be taken into account, too. { It should also be mentioned that the dwarf galaxies are also ideal candidates to constrain the annihilation cross section of dark matter. Geringer-Sameth \& Koushiappas (2012) performed a joint analysis of dwarf galaxy data and found that the upper limit on the annihilation cross section to a two-photon final state is $3.9^{+7.1}_{-3.7} \times 10^{-26}~{\rm cm^3~s^{-1}}$ at 130 GeV (see also Huang et al. 2012), which is well above that needed to account for the signal identified in the Galactic center (see below). }

\subsection{Fermi LAT data Analysis}
In last subsection we have suggested that $\psi \in (20^\circ,~40^\circ)$ excluding $b\leq 10^\circ$ may be a suitable region to constrain/measure $\langle\sigma v\rangle_{\rm \chi\chi \rightarrow \gamma\gamma}$. In reality, for a given dark matter density profile the intrinsic $\langle\sigma v\rangle_{\rm \chi\chi \rightarrow \gamma\gamma}$ should be smaller than the values or upper limits inferred from any other regions. Currently the tentative $\gamma-$ray signal is present in a very compact region, hence we are only able to set an upper limit on $\langle \sigma v \rangle_{\chi\chi\rightarrow \gamma \gamma}$ and impose a constraint on the slope of the dark matter density profile.

For such a purpose, we analyze the publicly available Fermi-LAT data in several regions: including (a) $\psi\leq 2^\circ$, which covers the most prominent signal region identified in \citet{tempel}; (b)  the region $\psi\in (2^\circ,~6^\circ)$; (c)  the region $\psi\in (6^\circ,~10^\circ);$(d) the region $\psi\in (10^\circ,~20^\circ)$ excluding $b\leq 10^\circ$; (e) the region $\psi \in (20^\circ,~30^\circ)$ excluding $b\leq 10^\circ$;(f) the region $\psi \in (30^\circ,~45^\circ)$ excluding $b\leq 10^\circ$; (g) the rest region $\psi \in (45^\circ, 180^\circ)$ excluding $b\leq 10^\circ$. We take into account the data in the time interval from 4 August 2008 to 18 April 2012 (MET 239557417 - MET 356439845), with energies between 20 and 200 GeV. We used the standard LAT analysis software (v9r27p1)\footnote{http://fermi.gsfc.nasa.gov/ssc}. To reduce the effect of the Earth albedo background, time intervals when the Earth was appreciably in field-of-view (FoV), specifically when the center of the FoV was more than $52^ \circ$ from zenith, as well as time intervals when parts of the ROI were observed at zenith angles $> 100^ \circ$ were also excluded from the analysis. The spectral analysis was performed based on the P7v6 version of post-launch instrument response functions (IRFs).  ULTRACLEAN dataset was selected to avoid the contamination from the charged particle. The spectra are shown in Fig.1.

\begin{figure}
\centering
\includegraphics[width=0.96\columnwidth,angle=0]{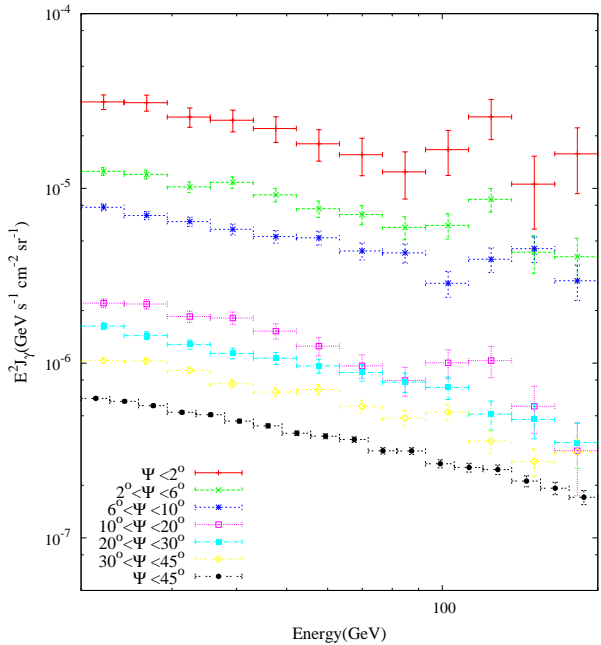}
\caption{The 20-200 GeV spectra of our Galaxy in seven regions described in the text (Please note that $b\leq 10^{0}$ is excluded for the last four regions). { $J_{\gamma}$ is the differential gamma ray flux. }} \label{fig:f1}
\end{figure}

To investigate  the line signal  at $130 ~\rm GeV$ we use the unbinned analysis method which is similar to the one described in \citep{fermiline}. It should be noted that in drawing fig.1 we binned the counts to different energy bins, which may introduce fake signal when the counts number is small. However, in fitting the data to derive the possible line signal or upper limits we use the unbinned analysis, say, the likelihood function was built by multiply the probability distribution function of each photons in the assuming model. This method can minimize the fake signal due to binning and take advantage of the full information of the observed data.   The likelihood is described as
\beq
{\cal L}=\prod_i f S(E_i)+(1-f) B(E_i),
\eeq
where $S(E_i)$ and $B(E_i)$ represent the signal and background function, respectively, both are normalized to 1, and $i$ runs over all the photons; $f$ is the signal fraction and has been set to be in the range $[-1, 1]$ for line signal search and $[0, 1]$ for getting upper limits; $B(E_i)$ takes the form
\beq
B(E_i) \sim E_i^{-\Gamma} \epsilon(E_i),
\eeq
where $\epsilon(E_i)$ is the exposure generated by the \emph{gtexpcube2} routine. $S(E_i)$ is derived by convolving the energy dispersion function and exposure.  Pyminuit \footnote{http://code.google.com/p/pyminuit.} are used to find the maximum of the likelihood. The MINOS asymmetric error at the level $\Delta \ln {\cal L} =1.35$ is adopted to get upper limit corresponding to a coverage probability of $95\%$.  To see the possible systematics due to the different choice of the background spectrum, we redo the analysis by using two different background spectral template $B(E_i)$: (1)  adding an exponential cutoff to the pure power law spectrum and leaving the cutoff  energy to be free, say,
\beq
B(E_i) \sim E_i^{-\Gamma} exp(-E_i/E_{cut})\epsilon(E_i),
\eeq
(2) a log-parabola spectrum
\beq
B(E_i) \sim (\frac{E_i}{E_b}) ^{-(\alpha+\beta log(E_i/E_b))}\epsilon(E_i).
\eeq
We do not find significant improvement of the fitting, thus for simplicity we use the pure power law spectrum as our fiducial background spectral model.

\begin{figure}
\centering
\includegraphics[width=0.96\columnwidth,angle=0]{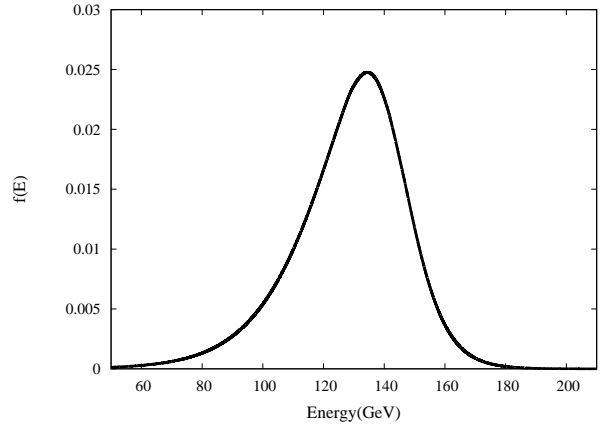}
\caption{The energy dispersion used in our analysis. The form of energy dispersion is described in the text and has already been normalized.
} \label{fig:edis}
\end{figure}

The region we choose is all larger than several degrees, meanwhile, the angular resolution above $20~\rm GeV$ is within 0.2 degree, which is much smaller than the size of the region of interest. Thus we neglect the point dispersion function(PSF) in the analysis. On the other hand, the energy dispersion is extremely important in the line searching process. In this work we focus on the tentative 130 GeV line, and adopt the energy dispersion at 130GeV with the form described on the website \footnote{http://fermi.gsfc.nasa.gov/ssc/data/analysis/documentation\\/Cicerone/Cicerone\_LAT\_IRFs/IRF\_E\_dispersion.html}.  The P7ULTRACLEANV6 version of the instrument response functions(IRFs) is used and the final energy dispersion is averaged for  different incidence angles.  { The derived energy dispersion used in the analysis is shown in Fig.\ref{fig:edis} and the signal flux (or upper limits) obtained in different regions are summarized in Tab.1.}

\begin{table}[htbp]
\caption{The signal flux as well as the $95\%$ confidence level upper limits obtained in different regions. Please note that $b\leq 10^{0}$ is excluded for the last four regions.} \label{tab:1} \centering
\begin{tabular}{cccc}
\hline
Region &Position& Signal/upper limit ($10^{-8}~\rm cm^{-2} s^{-1} sr ^{-1}$) \\

\hline

a   &$\psi\leq 2^\circ$& $3.8 \pm 1.6 $\\
\hline
b  &$\psi\in (2^\circ,~6^\circ)$&0.55\\

\hline
c   &$\psi\in (6^\circ,~10^\circ)$&0.34\\

\hline
d  &$\psi\in (10^\circ,~20^\circ)$&0.133\\

\hline
e   &$\psi\in (20^\circ,~30^\circ)$&0.028\\

\hline
f  &$\psi\in (30^\circ,~45^\circ)$&0.0084\\

\hline
g    &$\psi\geq 45^\circ$&0.0015\\
\hline
\end{tabular}
\end{table}

\subsection{Constraints on $\langle\sigma v \rangle_{\chi\chi\rightarrow \gamma \gamma}$ and the dark matter density profile}

To constrain dark matter distribution in inner Galaxy, we implement spherically symmetric generalized NFW profiles
\begin{equation}
\label{profile}
\rho_{_{\rm DM}(r)}=\rho_{\rm s}(r/r_{\rm s})^{-\alpha}(1+r/r_{\rm s})^{-3+\alpha},
\end{equation}
where $r_{\rm s}$ restricted in the range $10-35$ kpc is the scale radius \citep{Iocco2011} and $\alpha$ is the inner slope
for the NFW profile (it is found in the N-body simulations that $0.9<\alpha<1.2$, however the baryon compression may give rise to $\alpha\sim 1.7$). One main reason for adopting the generalized
NFW profiles is that they can reproduce the three types of dark halo profiles presented in Maccio et al. (2012).
The normalization of the DM profile is set by
the local DM density, i.e., $\rho_0 \equiv \rho_{_{\rm DM}}(R_\odot)$ and $\rho_{\rm s}=\rho_0(r_\odot/r_{\rm s})^{\alpha}(1+r_\odot/r_{\rm s})^{3-\alpha}$, where $R_\odot\approx 8.5$ kpc is the distance from the Sun to the Galactic center. The fiducial interval $\rho_0 = 0.4 \pm 0.1 ~{\rm GeV~cm^{-3}}$ is adopted in line with recent astrophysical measurements (Salucci et al. 2010).
The $\gamma$-ray flux produced by
DM annihilation can be written as
\begin{equation}
\label{eqn:phi}
\Phi(\Delta\Omega,E_{\gamma})=\frac{1}{4\pi} \times \frac{\langle
\sigma v\rangle_{\chi\chi\rightarrow \gamma\gamma}}{2m^2_{\chi}}\,\frac{\mathrm{d}N_{\gamma}}{\mathrm{d}
E_{\gamma}} \times \bar{J}(\Delta\Omega)\Delta\Omega,
\end{equation}
where $m_{\chi}$ is the mass of DM particles and
${{\rm d}N_{\gamma}/{\rm d}E_{\gamma}}=2\delta(E_\gamma-m_\chi)$ is the differential energy spectrum of $\gamma$-rays. The astrophysical factor ($\bar{J}$) is
defined as
\begin{equation}
\label{eqn:jbar}
\bar{J}(\Delta\Omega) = \frac{1}{\Delta\Omega}
\int_{\Delta\Omega}{\rm d}\Omega\int_{\rm LOS}{\rm d}l\,\rho^2(r(l)),
\end{equation}
where $r(l)$ is the distance to the
center of the object which is a function of line of sight (LOS) distance $l$.

\begin{figure}
\centering
\includegraphics[width=0.96\columnwidth,angle=0]{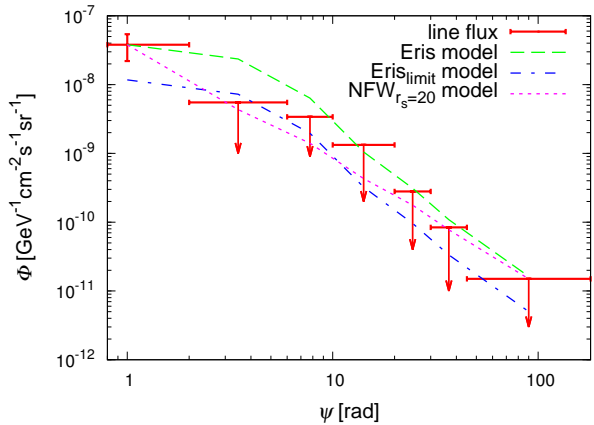}
\caption{The annihilation line flux expected in the theoretical models vs. the current 130 GeV line data. See the text for details.} \label{fig:f3}
\end{figure}

For region (a), there is a tentative gamma-ray line signal. If interpreted as
dark matter particles annihilating into a pair of photons, the cross section of dark matter
annihilation ($\langle\sigma v\rangle_{\rm {\chi\chi\rightarrow \gamma\gamma}, a}$) can be inferred with equation (\ref{eqn:phi}). For regions (b), (c), (d), (e), (f) and (g) we have the $95\%$ confidence level upper
limits of the line flux and then get the constraints on the annihilation cross section. As long as
\begin{eqnarray}
\langle\sigma v\rangle_{{\chi\chi\rightarrow \gamma\gamma},\rm a} &\leq & \min\{ \langle\sigma v\rangle_{{\chi\chi\rightarrow \gamma\gamma}, \rm b}, ~\langle\sigma v\rangle_{{\chi\chi\rightarrow \gamma\gamma}, \rm c}, \nonumber\\
&& ~\langle\sigma v\rangle_{{\chi\chi\rightarrow \gamma\gamma}, \rm d},~\langle\sigma v\rangle_{{\chi\chi\rightarrow \gamma\gamma}, \rm e}, \nonumber\\
&&
~\langle\sigma v\rangle_{{\chi\chi\rightarrow \gamma\gamma}, \rm f},~\langle\sigma v\rangle_{{\chi\chi\rightarrow \gamma\gamma}, \rm g}\},
\end{eqnarray}
the Dark Matter profile is in agreement with the 130 GeV $\gamma-$ray
line data. Such a fit to the current tentative 130 GeV $\gamma-$ray line data suggests that $\alpha
\geq 1.17$ and $\langle\sigma v\rangle_{\chi\chi\rightarrow \gamma\gamma} \leq 7.5 \times 10^{-28}~{\rm cm^3~s^{-1}}$ for $r_{\rm s}=20$ kpc and $\rho_0=0.4~{\rm GeV~cm^{-3}}$ (see the dotted line in Fig.\ref{fig:f3}).

\begin{figure}
\centering
\includegraphics[width=0.9\columnwidth,angle=0]{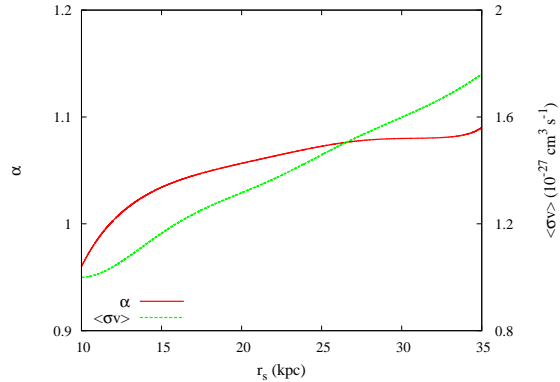}
\caption{ The 95\% confidence level lower (upper) limit on the generalized NFW distribution parameter $\alpha$ ($\langle \sigma v \rangle_{\chi\chi\rightarrow \gamma\gamma}$) set by current 130 GeV line data for $\rho_0=0.4~{\rm GeV~cm^{-3}}$. Following \cite{Iocco2011}, $r_{\rm s}$ is restricted within the range $10-35$ kpc.}  \label{fig:f4}
\end{figure}

{ To get more robust constraint, we adopt a combined likelihood analysis in all seven regions to constrain $\langle\sigma v\rangle_{\chi\chi\rightarrow \gamma\gamma}$ and $\alpha$ for $r_{\rm s}$ ranging from 10 kpc to 35 kpc. The method is similar to that described in \citet{tsai12} and the combined likelihood is calculated as
\begin{equation}
L_c=\Pi L_i,
\end{equation}
where $L_i$ is the unbinned likelihood in the $i-$th region. The definition of the unbinned likelihood is similar to that in last section, but now the influence of $\langle\sigma v\rangle_{\chi\chi\rightarrow \gamma\gamma}$ as well as $\alpha$ should be taken into account. For such a purpose we modify the signal ratio $f$ in Eq.(1) to $f=\Phi(\langle\sigma v\rangle_{\chi\chi\rightarrow \gamma\gamma}, \alpha)/\Phi_{\rm obs}$, where $\Phi(\langle\sigma v\rangle_{\chi\chi\rightarrow \gamma\gamma}, \alpha)$ is the flux predicted in the DM model (i.e., eq.(\ref{eqn:phi})) with varying $\langle\sigma v\rangle_{\chi\chi\rightarrow \gamma\gamma}$ and $\alpha$,
and $\Phi_{\rm obs}$ is the observed gamma ray flux (i.e., the integration of $J_{\gamma}$ in Fig.1 in the whole energy range).}  In our approach the profile likelihood technique is adopted \citep{rolke05}. To constrain $\langle\sigma v\rangle_{\chi\chi\rightarrow \gamma\gamma}$ we treat $\alpha$ as a nuisance parameter and vice versa. $\Delta \ln(L_c) = 1.35$ is adopted to get upper (lower) limit corresponding to a coverage probability of 95\%. In this work we do not constrain $\rho_0$ (or alternatively $\rho_{\rm s}$) since it couples with the annihilation cross section, i.e., $\langle\sigma v\rangle_{\chi\chi\rightarrow \gamma\gamma} \varpropto 1 / \rho_{\rm 0}^2$. Our general 95\% confidence level lower (upper) limit on $\alpha$ ($\langle\sigma v\rangle_{\chi\chi\rightarrow \gamma\gamma}$) as a function of $r_{\rm s}$ is presented in Fig.\ref{fig:f4}. For $r_{\rm s}=20$ kpc and $\rho_0=0.4~{\rm GeV~cm^{-3}}$, the 95\% confidence level constraints are $\alpha \geq 1.06$ and $\langle\sigma v\rangle_{\chi\chi\rightarrow \gamma\gamma} \leq 1.3 \times 10^{-27}~{\rm cm^3~s^{-1}}$ , respectively.  The required $\langle\sigma v\rangle_{\chi\chi\rightarrow \gamma\gamma}$ is consistent with the constraints set by the non-detection of a reliable signal in dwarf galaxies, diffuse Galactic halo and Galaxy Clusters \citep{strigari12,bringman2012}. Even for $r_{\rm s}=10$ kpc, the smallest value suggested in \cite{Iocco2011},
$\alpha\geq 0.96$ is needed and hence the HFR dark matter profile
suggested in Maccio et al. (2012) is disfavored,
implying that the baryonic processes that can considerably flatten the central
dark matter distribution might not play an important role in the
evolution of the Milky Way.

In Eris, a very recent high-resolution cosmological
hydrodynamics simulation of a realistic Milky-Way-analog disk galaxy, the dark matter density profile within $\sim 1$ kpc is found to be very flat and the peak of the dark matter density profile is $\sim$ a few 100 pc away from the Galactic centre (Kuhlen et al. 2013). It is interesting to check whether the specific cored dark matter distribution found in Eris is consistent with the current line data or not. The fits to the data are shown in Fig.\ref{fig:f3}. The dashed line represents the best fit. In region (b) the divergency between the model and the data is by a factor of $>4$. The dash-dotted line (i.e., the so-called Eris$_{\rm limit}$ model) represents the case that the predicted flux of Eris model in region (a) has been normalized to a flux $\approx 1.17\times 10^{-8}~{\rm cm^{-2} s^{-1} sr ^{-1}}$, i.e., the 95\% confidence level lower limit of the line signal. In such a fit, the model predicted line flux in region (b) is still above the upper limit. We thus conclude that the Eris model has some tension with the line data.

\section{Summary}
In the very inner region of the Galaxy, the dark matter abundance is much less than the normal matter, hence current micro-lensing and Galactic rotation curve data can not directly constrain the dark matter density distribution. The numerical simulations in principle can solve such a problem. However, our current knowledge of the history of the Galaxy is very limited and the complicated physical processes being able to shape the dark matter distribution are hard to fully address. That's why so far the distribution of dark matter in the very inner region of the Galaxy is still in heavy debate (e.g., Gnedin et al. 2004; Maccio et al. 2012; Pontzen \& Governato 2012; Kuhlen et al. 2013). Recently several groups have identified tentative 130 GeV $\gamma-$ray line which might be due to the annihilation of dark matter particles in the inner Galaxy. { In this work we adopt the hypothesis that these signals are due to dark matter annihilation and use this hypothesis to examine which DM profile is consistent with the line. Our finding is that at the 95\% confidence level, the dark matter density profile towards the center should be not shallower than $r^{-1.06}$ (for the generalized NFW profile with $r_{\rm s}=20$ kpc) and the dark matter annihilation cross section should be smaller than $\langle \sigma v \rangle_{\chi\chi\rightarrow \gamma\gamma} =1.3\times 10^{-27}~{\rm cm^{3}~s^{-1}}(\rho_0/0.4~{\rm GeV~cm^{-3}})^{-2}$ (see Fig.4).} Such a density profile is in agreement with that found in some N-body simulations and the baryon compression effect might play a role. The dissipative baryonic processes that are able to considerably flatten dark matter profiles seems to not play important roles in our Galaxy, implying that the formation and evolution of Milky Way may not follow a prototypical spiral galaxy. Finally we'd like to caution that the tentative 130 GeV $\gamma-$ray line has not been officially confirmed by the Fermi collaboration yet. Whether our constraints on the dark matter distribution in very inner Galaxy and on the corresponding annihilation cross section are robust or not will be directly tested by the upcoming pass 8 data of Fermi LAT, in which the amount of usual data at energies greater than 10 GeV is expected to be boosted by some $60\%$ \citep{pass8}.

\acknowledgments
We are grateful to the anonymous referee for the insightful comments
that help us to improve the manuscript significantly. We also thank Dr. Qiang Yuan for helpful suggestion. This work is
supported in part by the 973 Program of China (No. 2013CB837000), 100
Talents program of Chinese Academy of Sciences, Foundation for Distinguished Young
Scholars of Jiangsu Province, China (No. BK2012047), and the China
Postdoctoral Science Foundation (No. 2012M521136). \\

\end{document}